\documentclass[fleqn,twoside]{article}
\usepackage{espcrc2}
\usepackage{graphicx}

\newcommand\fig[1]{Fig.~\ref{#1}}
\newcommand\eq[1] {Eq.~\ref{#1}}
\newlength{\www}
\newcommand {\EPSFigure}[4] {
  \begin{figure}
  \begin{center}
  \includegraphics[width=#2\www]{#1}
  \end{center}
  \vskip-10mm                          
  \caption{#4}
  \vskip-8mm                           
  \label{#3}
  \end{figure}
}

\def\ns   {\mathrm{\,   ns}}

\def\uV   {\mathrm{\,\mu V}}
\def\V    {\mathrm{\,    V}}
\def\um   {\mathrm{\,\mu m}}
\def\uA   {\mathrm{\,\mu A}}
\def\pA   {\mathrm{\,  p A}}

\def\kOhm {\mathrm{\,  k\Omega}}

\def\e    {\mathrm{\,  e^-}}
\def\bit  {\mathrm{\,bit  }}
\def\SA   {\mathrm{S_1}    }
\def\SB   {\mathrm{S_2}    }
\def\SC   {\mathrm{S_3}    }
\def\Iin  {\mathrm{I_{in} }}
\def\Iout {\mathrm{I_{out}}}
\def\IO   {\mathrm{I_{0}}}
\def\Ifix {\mathrm{I_{fix}}}
\def\ID   {\mathrm{I_{D}  }}
\def\gq   {\mathrm{g_{q}  }}
\def\gm   {\mathrm{g_{m}  }}
\def\Vgs  {\mathrm{V_{gs} }}
\def\Qin  {\mathrm{Q_{in} }}
\def\Cgd  {\mathrm{C_{gd} }}
\def\Cgs  {\mathrm{C_{gs} }}
\def\CL   {\mathrm{C_{L}  }}

\title{Readout Concepts for DEPFET Pixel Arrays}

\author{
  P.~Fischer\thanks{Corresponding author. Tel.: +49-228-732996; fax: +49-228-733220.
                    E-mail address: fischerp@physik.uni-bonn.de},
  W.~Neeser,
  M.~Trimpl,
  J.~Ulrici,
  N.~Wermes\address[]{Physikalisches Institut der Universit\"at Bonn, Germany}
}

\begin{document}

\begin{abstract}
Field effect transistors embedded into a depleted silicon bulk (DEPFETs) can be used as the first amplifying element
for the detection of small signal charges deposited in the bulk by ionizing particles, X-ray photons or visible light.
Very good noise performance at room temperature due to the low capacitance of the collecting electrode has been
demonstrated. Regular two dimensional arrangements of DEPFETs can be read out by turning on individual rows and reading
currents or voltages in the columns. Such arrangements allow the fast, low power readout of larger arrays with the
possibility of random access to selected pixels. In this paper, different readout concepts are discussed as they are
required for arrays with incomplete or complete clear and for readout at the source or the drain. Examples of VLSI
chips for the steering of the gate and clear rows and for reading out the columns are presented.
\\
\\
\noindent Keywords: DEPFET, monolithic active pixel sensors, integrated amplification, readout electronics
\\
\\
\noindent
PACS: 06.30.Bp, 
      07.50.Ek, 
      29.40.Gx, 
      87.58.Mj, 
      87.59.Bh  
\end{abstract}

\maketitle

\section{Introduction} \label{SECT_INTRO}

Silicon sensors are widely used for the detection of visible photons, soft x-rays and ionizing particles in many
industrial and medical applications as well as in particle physics. An important figure of merit is the noise of the
sensor. It should be as low as possible to improve the spectroscopic performance, to increase the spatial resolution
and to be able to detect low charge depositions for instance from low energetic $\gamma$-rays (e.g. $^3T$) or from
minimum ionizing particles in thin sensors ($<50\um$) as they are required in the innermost layers of vertex detectors
for particle physics \cite{TESLA_PROPOSAL}. A very low noise of $5\e$ rms has been demonstrated \cite{DEPFET_NOISE}
with single DEPFET devices which integrate a p-channel field effect transistor into a fully depleted bulk
\cite{DEPFET_ORGINAL}. Electron-hole pairs generated by photon absorption or by ionizing radiation are separated in the
electric field of a sidewards-depleted bulk. The electrons drift to a region under the transistor channel, the {\em
internal gate}, where they modulate the drain current of the FET according to $\Delta \ID = \gq \Delta Q$ for small
charges. In addition to the modulation through the internal gate, the drain current can be switched off by a positive
potential at the {\em external gate}. This makes it possible to arrange several square, rectangular or hexagonal DEPFET
devices in an array made out of columns and rows as explained in the next section.

The charges accumulated in the internal gate remain in the local potential minimum so that multiple non-destructive
readout is possible. They must be removed regularly to avoid saturation and to prepare a well defined initial
condition. The {\em clearing} of charges from the internal gate can be achieved with positive signals on separate clear
contacts. The design of these structures is difficult because no charges should be lost into the clear contacts in
normal operation mode while all charges should be removed from the internal gate while clearing. A complete clearing
could not be achieved on existing DEPFET devices even with voltages of up to $20\V$ so that the fill state of the
internal gate has to be measured and memorized after the clear pulse by suited readout electronics. A new generation of
DEPFET devices \cite{DEPFET_HOLL} should provide a complete clear so that different readout concepts become possible.

After an introduction of the matrix arrangement and the time sequence of the readout, we describe existing readout
electronics for devices with incomplete clear. We then propose a concept for the fast drain readout of devices with
complete clear.

\section{Readout of DEPFET arrays}

\EPSFigure {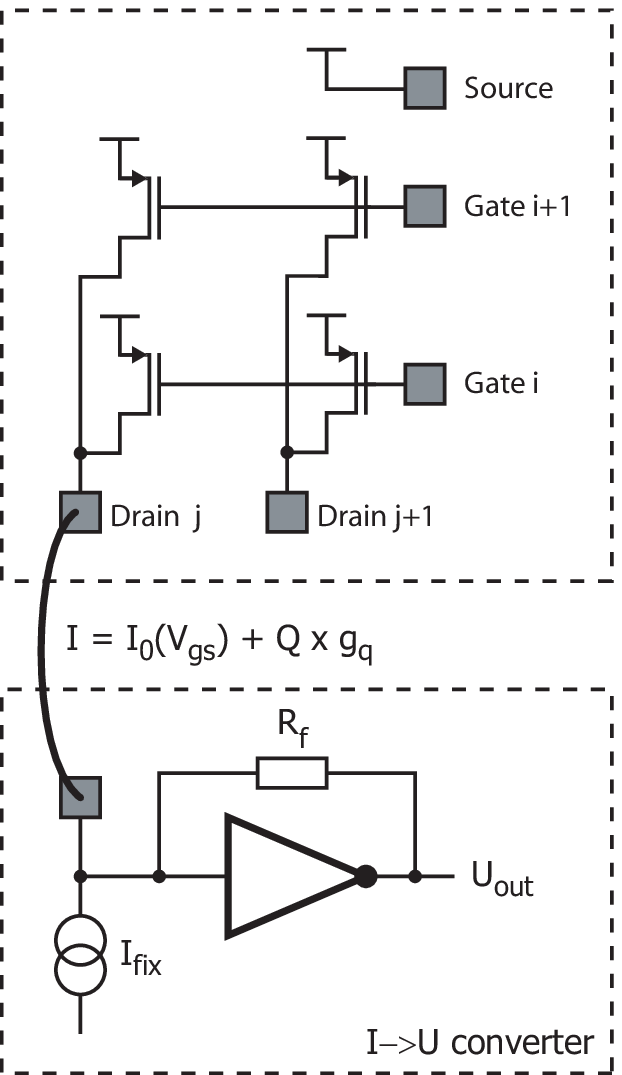}{0.7}{FIG_MATRIXREADOUT} {Two dimensional DEPFET array with readout at the
drain}

The arrangement of many DEPFET devices in a two dimensional array is schematically shown in \fig{FIG_MATRIXREADOUT} for
the case of a current readout at the drain. The external gates and the clear contacts (not shown for simplicity) are
connected horizontally in rows while the drains are connected in columns. The drain current in all devices except in
one row is switched off by applying a positive voltage to the external gates. The external gates of the p-channel
DEPFETs in the active row are connected to a more negative voltage to enable a current flow in the drain. This current,
which is accessible at the column contacts, depends on $\Vgs$ and on the charge accumulated in the internal gate. Any
noise on the external gate is amplified by the transconductance of the transistor so that a low noise voltage source is
required. This problem is avoided by connecting the gate row directly to source potential so that $\Vgs=0$. The drain
current is then mainly determined by the geometry and doping profiles of the DEPFET devices. The drain current is often
converted to a voltage for instance with the simple $U\rightarrow I$ converter sketched in \fig{FIG_MATRIXREADOUT}. The
offset current $\IO$, which is not carrying any information, can be subtracted immediately by a constant current source
$\Ifix$ so that the dynamic range of the converter is not wasted.

\subsection{Readout sequences for incomplete and complete clear}

\EPSFigure {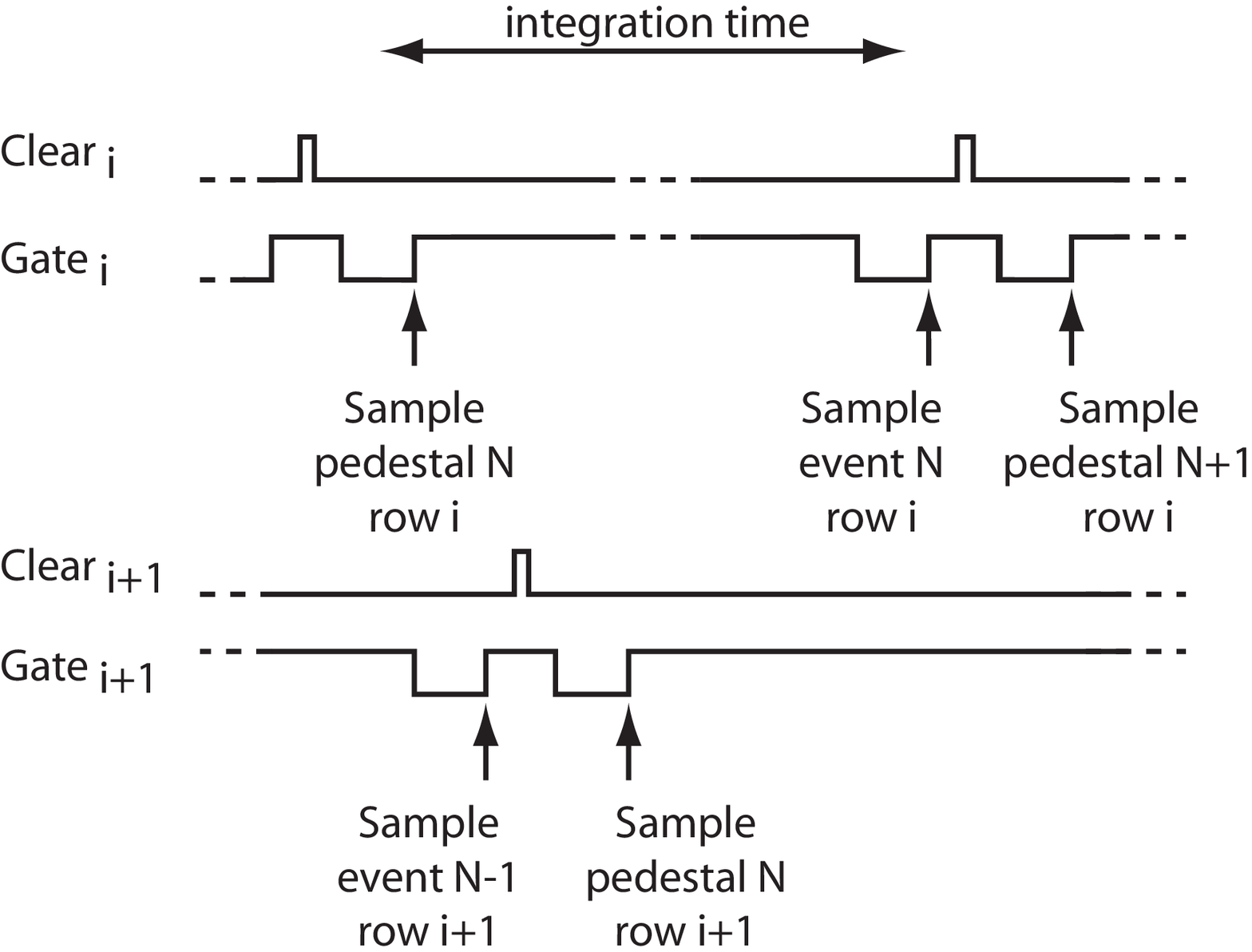}{1.0}{FIG_Timing2}{Control signal sequence for a DEPFET array with incomplete
clear. The pedestal values for every frame must be measured after the clear pulse}

The switching sequence of the various gate and clear rows and the processing of the currents at the bottom of the
columns depends very much on the characteristics of the clearing procedure.

For devices with {\em incomplete clear}, the charge left in the internal gates must be determined after every clear
pulse by turning on the row as indicated in \fig{FIG_Timing2}. The row is then re-measured after the frame interval.
The charge accumulated during the integration time in the internal gate is calculated as the difference of these two
drain currents divided by the $\gq$ of the device. The other rows must be processed between the pedestal- and the
signal measurement so that many different currents are present in a column before the signal measurement can be done.
The pedestal values of the complete array must therefore be memorized during the frame interval with an absolute
precision of less than the targeted pixel noise. The required relative precision is very high due to the large standing
pedestal current in the DEPFET device. For a current of $100\uA$ and devices with $\gq=200\pA/\e$ for instance, a
relative precision of $2\times 10^{-5}$ is required for a fluctuation of $10\e$. Note that this requirement concerns
mainly the noise introduced in the sampling procedure while systematic errors (for instance due to charge injection)
cancel out in the subtraction. The requirement can be relaxed by an order of magnitude if a large fraction ($\approx
90\%$) of the pedestal current is subtracted with a low noise current source prior to the $U\rightarrow I$ conversion
as shown in~\fig{FIG_MATRIXREADOUT}. The temporary storage of the pedestal values could be done on a readout chip at
the bottom of {\em small} arrays, but it is not feasible for large structures. The data must therefore be processed by
the full readout system which must provide a large dynamic range to cope with fluctuations of the standing drain
current in the order of $5-10\%$ for existing arrays.

\EPSFigure {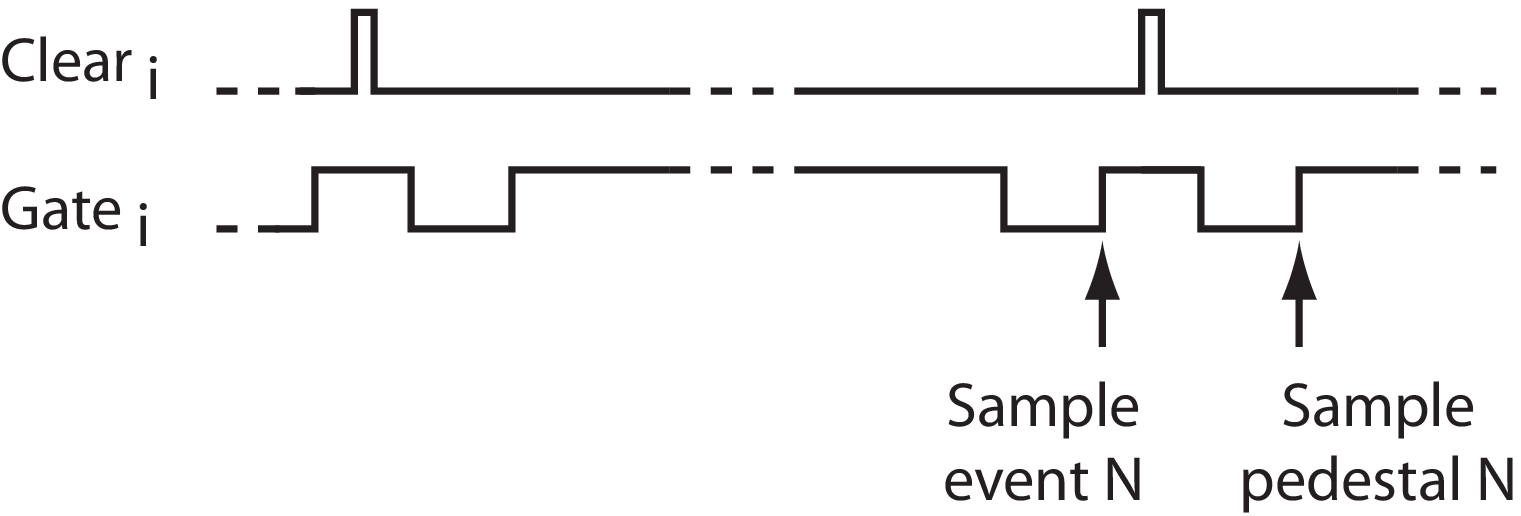}{1.0}{FIG_Timing1}{Control signal sequence for a DEPFET array with
complete clear. The pedestal values for a frame can be measured after the next clear}

For devices with {\em complete clear}, a unique pedestal value per pixel measured just after the clear pulses for
instance at the beginning of the run could be used. This approach would, however, still require the measurements of
large signals and an 'off-line' subtraction. The readout sequence shown in \fig{FIG_Timing1} avoids this by comparing
the signal current with the pedestal value after the {\em next} clear pulse. Note that this is only possible because
the complete clearing leaves the internal gate in the same well defined (empty) state. Because signal and pedestal
measurements are following each other immediately, the subtraction can be easily done on the readout chip and no long
time storage of large signals with high precision is required. The noise performance could be degraded in this scheme
if the clear pulse introduces additional noise contributions, for instance by changing the state of traps in the
transistor channel. This must be further studied in the future.

\subsection{Drain and source readout}

\EPSFigure {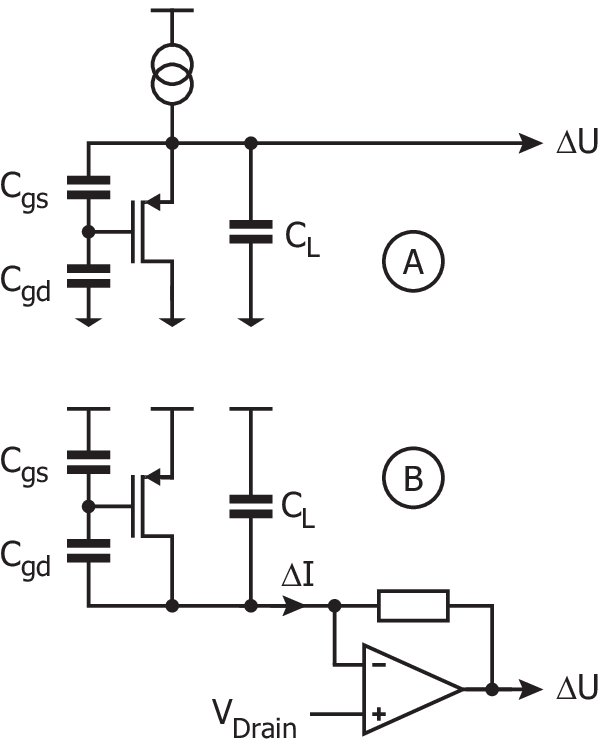}{0.8}{FIG_DrainSource}{Simplified equivalent circuit diagrams for readout
at the source (A) or at the drain (B)}

The vertical busses of the array can be connected to the drains or to the sources of the individual DEPFET devices. In
the practical design of the array, however, other considerations can impose one or the other type of readout. (If the
vertical bus is, for instance, implemented by connecting the outer implants of the individual DEPFETs, the more
negative potential must be used there to push the electrons towards the internal gate at the inside of the annular
structure. The bus must therefore be connected to the drains in this case.) The two readout types are sketched in the
simplifying schematic of \fig{FIG_DrainSource} which shows the DEPFET together with the parasitic capacitances from
drain and source to the internal gate.

For readout at the {\em source} (\fig{FIG_DrainSource}A), a current is forced through the device by an external current
source. The voltage change at the source in this follower configuration is measured. It is given approximatively by
\cite{DEPFET_RISETIME}
\begin{equation}
  \Delta U \approx {\Qin \over \Cgd} \label{EQ_SFVOLT}
\end{equation}
as long as the transconductance of the device is large. $\Cgs$ does not play a role in this expression because $\Vgs$
remains constant due to the unity gain of the arrangement. The rise time of the voltage signal is severely degraded by
the bus capacitance $\CL$ which must be charged by the current change in the DEPFET. This current change is small,
however, due to $\Vgs$ being nearly constant. One finds \cite{DEPFET_RISETIME}
\begin{equation}
 t_r = 2.2 { \CL \left( 1+{\Cgs\over\Cgd}\right) +\Cgs \over \gm}.
\end{equation}
This settling time constant can easily reach several microseconds so that the source follower readout is not suited for
very high speed applications.

For readout at the {\em drain}, the charge in the internal gate leads to a current change which can be converted to a
voltage by a simple $I\rightarrow U$ converter as indicated in \fig{FIG_DrainSource}B. The converter can hold the bus
at a constant potential so that the bus capacitance $\CL$ does not affect the speed anymore. The gain is given by
\begin{equation}
 \Delta I \approx{\gm\over\Cgs+\Cgd}\, \Qin.
\end{equation}
A drawback of this readout scheme is the sensitivity to fluctuations in the device threshold and to voltage drops on
the source traces which are amplified by the DEPFET so that fairly large current fluctuations from pixel to pixel must
be coped with.

\section{Examples of chips for steering and readout}

The central elements of several different chips suited for the readout of DEPFET arrays are presented in this section.
As discussed above, a distinction between readout at source/drain and for devices with complete/incomplete clearing of
the internal gate can be made.

\subsection{Control of gate and clear voltages}

\EPSFigure {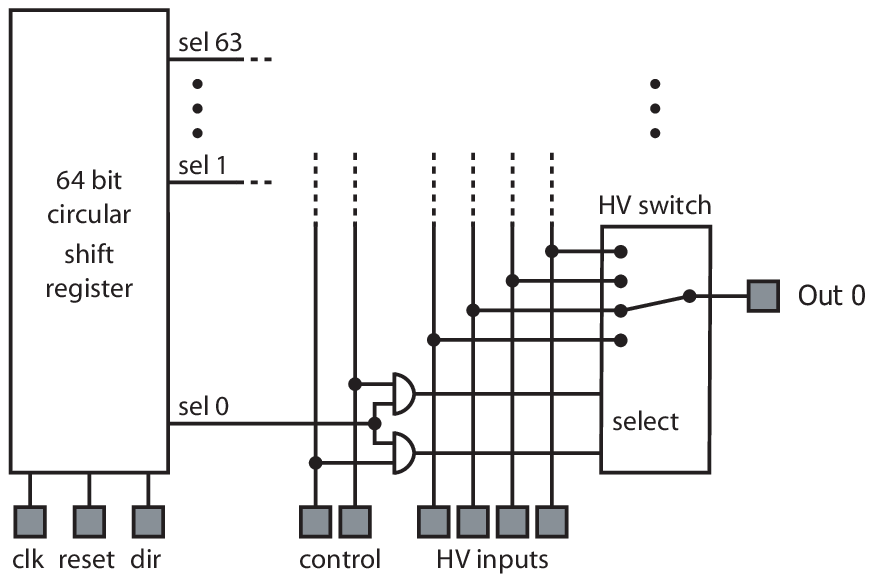}{1.0}{FIG_SWITCHER}{Simplified block diagram of the 'SWITCHER' gate/clear control
chip}

The gates of the DEPFET rows must be activated consecutively during the readout cycle. Furthermore, the row-wise clear
signals and, in more advanced DEPFET structures, a clear-control signal must be driven. The voltage swings of these
signals of up to $20\V$ exceed the supply voltages of 'standard' CMOS technologies, so that a 'high voltage' technology
with an allowed supply range of above $20\V$ has been used for the implementation of the simple 'SWITCHER' gate/clear
steering chip. As indicated in \fig{FIG_SWITCHER}, it consists of a shift register to select one out of 64 channels and
of analog switches to connect one of 4 voltages to the selected output. The shift register can be operated in both
directions. This is required if two chips are used on opposite sides of a DEPFET array for the gates and the clear
signals. The shift register can be switched to 'loop'-mode after initialization so that no further control signals are
required. The analog switches are implemented with paralleled NMOS and PMOS devices to keep the resistance low for all
voltages. A low switch resistance is important to allow for fast settling of the gate voltage and to keep the noise on
the sensitive gates low. (This also requires very low noise power supplies for the gate voltages.) Reducing the switch
resistance by using wider devices adds a significant capacitive load to the gate rows which slows down the setting
unnecessarily so that a compromise must be found. The present SWITCHER chip has 4 available voltages to allow for more
complicated switching sequences for detailed studies of the DEPFET characteristics in the array. A next generation
SWITCHER chip with increased functionality (on-chip biases, simple multi-chip operation, internal sequencer) is
presently being developed.

\subsection{Drain readout for DEPFETs with incomplete clear}

\EPSFigure {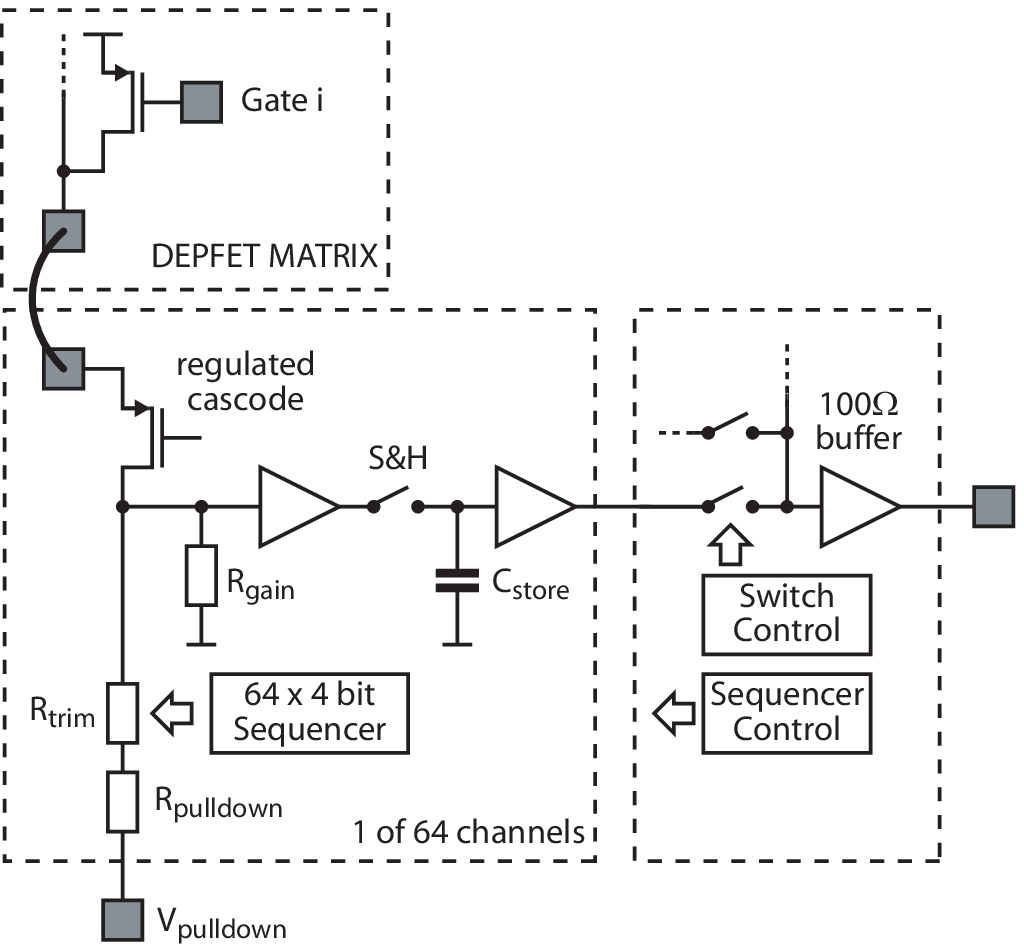}{1.0}{FIG_CARLOS}{Block diagram of a chip for drain readout of DEPFET arrays with
incomplete clear}

The block diagram of the first drain readout chip for DEPFET arrays with incomplete clear is sketched in
\fig{FIG_CARLOS}. The drain current, which is fixed by the voltage of the gate rows, is converted into a voltage and
sent to an ADC. The subtraction of the currents just after the clearing of the internal gate and at the end of the
frame is done off-line using the ADC values. As motivated above, the voltages at the drains are kept nearly constant by
means of a regulated cascode circuit. A constant current is subtracted from the drain current by a $60\kOhm$
polysilicon resistor connected to the negative pull-down voltage. This voltage can be set well below the chip ground
because no other components are connected to this node and no input protection is used. A large value for the pull-down
voltage ($-15\V$) and for the resistor was chosen to reduce its thermal noise below the thermal noise in the DEPFET
device. The remaining current is converted into a voltage simply by means of a large value resistor. The value of this
resistor is a compromise between high gain (reducing noise contributions of following stages) and dynamic range.
Measurements of fabricated DEPJFET arrays showed that the drain currents for a fixed gate voltage varied by more than
$10\%$. This is due to inhomogeneities in the doping profiles and due to IR-voltage drops on source and drain signals.
An individual adjustment of the subtracted current was therefore implemented to be able to keep the gain resistor high.
The digital correction values are stored in a $64\times 64\times 4\bit$ on-chip memory. They are used to increase or
decrease the value of the pull-down resistor in 16 steps. Using this readout chip, a DEPFET array can be operated fully
read out within the dynamic range at a gain of $60 \uV/\e$. The chip has been successfully used in the 'Bioscope'
system to operate a $64\times 64$ matrix of DEPJFETs \cite{DEPFET_BIOSCOPE} with an intrinsic speed of up to 1000
frames per second.

\subsection{Source follower readout for DEPFETs with complete clear}

\EPSFigure {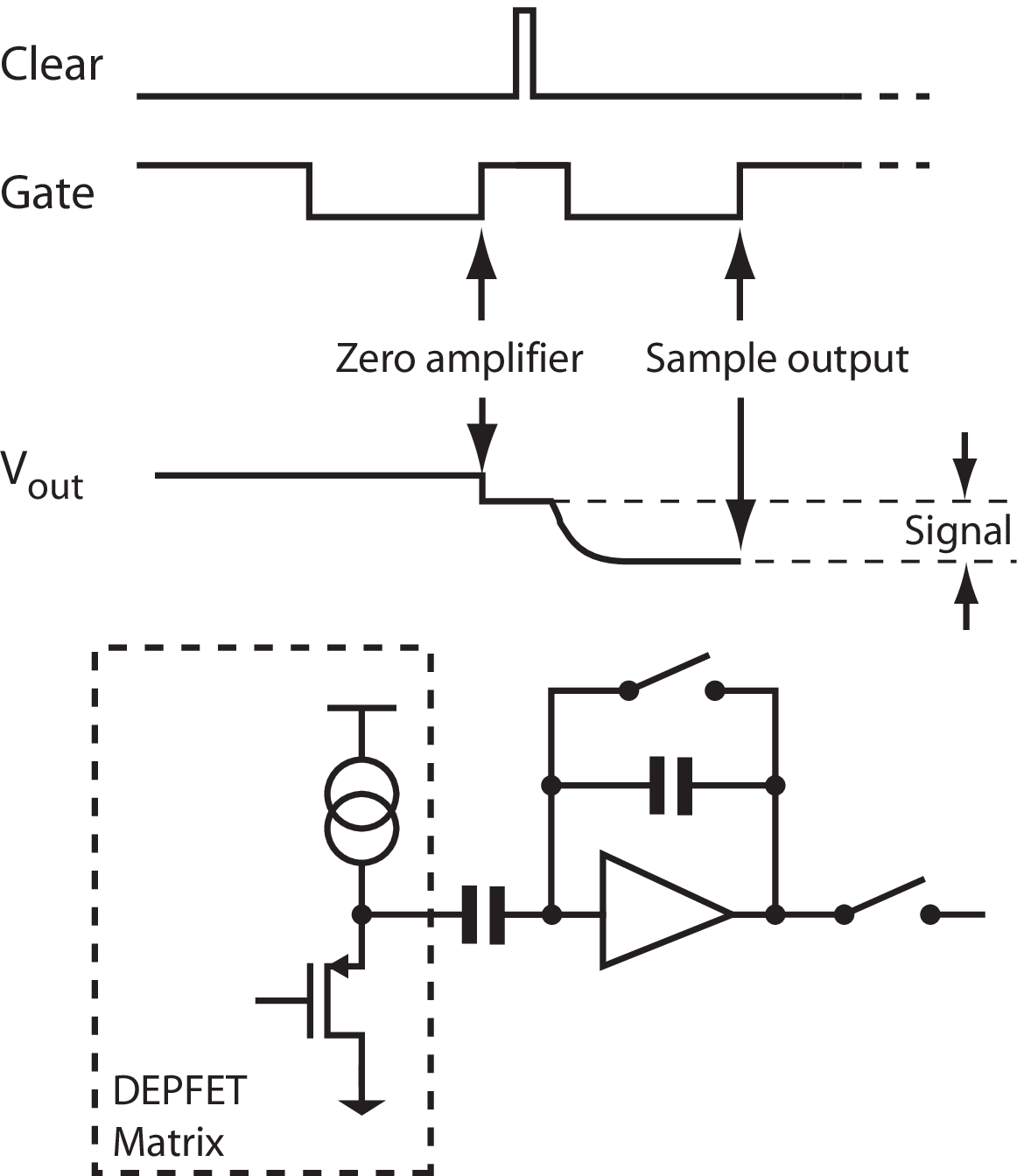}{0.8}{FIG_SFREADOUT}{Source follower readout of DEPFET array with complete clear}

The source readout of a matrix with complete clear can be achieved by a voltage amplifier with a reset switch. As
indicated in \fig{FIG_SFREADOUT}, the voltage amplifier (in this case implemented with a capacitive feedback) is set to
a known state while the activated DEPFET row outputs the 'signal' voltage. The change in the DEPFET output voltage
after clearing the row (according to \eq{EQ_SFVOLT}) is then amplified by the gain of the voltage amplifier. Note that
large spikes on the source bus during the application of the clear pulse could drive the amplifier into saturation and
degrade the output signal.

\subsection{Fast drain readout for DEPFETs with complete clear}

The block diagram of a chip for a fast drain readout assuming complete clear is sketched in \fig{FIG_TESLA}. A
regulated cascode circuit in every channel is used to keep the DEPFET drains at constant potential in order to decrease
the influence of the drain bus capacitance. After selecting a DEPFET row for readout, the signal current is stored in a
fast current memory cell (see below). The row is then cleared and the pedestal currents are subtracted from the stored
signal current simply by summing currents (the current memory cell outputs the negative of its input current). A
simple, fast current comparator is used to find the interesting hits, i.e. channels with a sufficiently large signal. A
neighbor logic (not shown) could also select channels in the vicinity of hits for readout. When there is at least one
hit in the row, the analog signal current and the binary hit flag of every channel are stored in a FIFO - like
structure which consists of current memory cells and digital storage bits. Only the currents which are stored in hit
cells are needed later so that all others cells can be turned off to save static power. The FIFO is thus filled
permanently with events containing at least one hit. It serves as a buffer to cope with statistical fluctuations in the
number of hits.

\EPSFigure {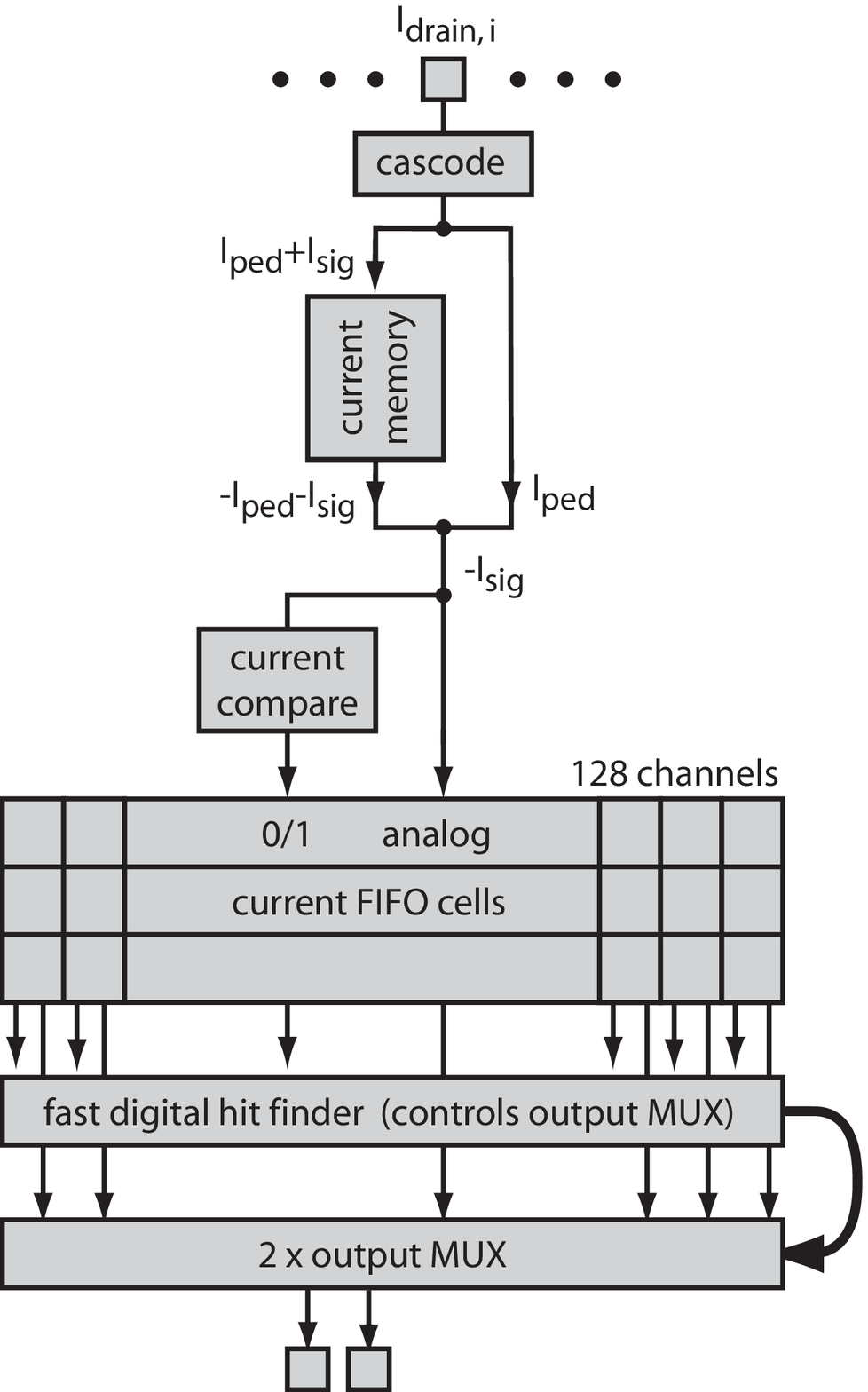}{0.9}{FIG_TESLA}{Block diagram of a chip for fast drain readout of DEPFET arrays with
complete clear}

The second part of the chip empties the FIFO row by row. A fast binary hit scanner \cite{MEPHISTO} can find up to two
hits per clock cycle. The selected current values are digitized on-chip or sent off-chip together with their addresses.
The scanner can process the FIFO row until no more hits are found so that there is no limitation on the number of hits
per row as long as the {\em average} hit rate can be coped with.

Several other required circuit elements (row counting, time stamping of events) are not shown in  \fig{FIG_TESLA}. The
proposed concept can be easily further improved if required. It is for instance possible to memorize the signal
currents during several clock cycles in a bank of memory cells at the input and perform the subtraction of the pedestal
current several clock cycles later. This could be advantageous if the clearing of the internal gate is not fast enough.

\EPSFigure {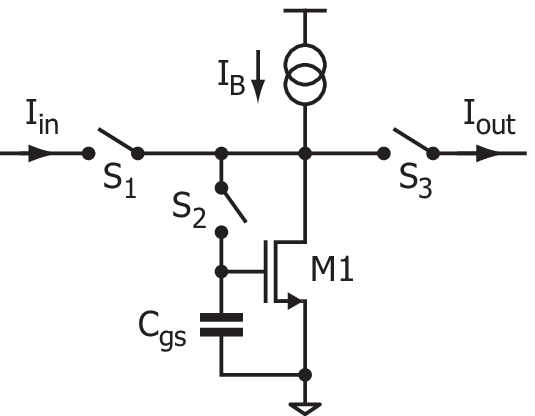}{0.8}{FIG_STROMSPEICHER}{Basic concept of the current memory cell used in the
readout chip}

One of the crucial building blocks in this concept is the current memory cell \cite{SWITCHEDCURRENT} used for pedestal
subtraction and for temporary analog storage in the FIFO-like structure. The basic concept of such cells is shown in
\fig{FIG_STROMSPEICHER}. When switches $\SA$ and $\SB$ are closed, the gate voltage of the NMOS transistor settles to
such a value that the sum of the input current $\Iin$ and the bias current $\mathrm{I_B}$ flow through the device. When
$\SB$ is opened, this voltage is stored on the (parasitic) gate-source capacitance $\mathrm{C_{gs}}$. The current
through the device remains unchanged ideally. $\SA$ can now be opened without affecting the stored gate voltage. The
current can be retrieved by closing $\SC$. In the ideal case one gets $\Iout = - \Iin$. This is however not perfectly
true due to charge injection onto $\mathrm{C_{gs}}$ when opening $\SB$ and due to the finite output conductance of the
storage device and the current source. The devices have therefore been cascoded and two stages have been used to cancel
charge injection. Preliminary measurements from a test chip designed in a $0.24\um$ technology using radiation tolerant
layout techniques \cite{RADHARD} have shown that shifting of currents is possible in less than $50\ns$ with a precision
of well below $1\%$.

\section{Summary}

The intrinsic advantages of DEPFET devices like low noise, thin entrance window and non-destructive readout can be
exploited on large areas by constructing DEPFET arrays. Gate and clear contacts are connected in rows while the drains
or sources are connected in columns. Several chips to steer the array and to read out the voltages or currents at the
bottom have been presented in this paper. The readout sequence is considerably simplified if the charge in the internal
gate can be completely removed with a suited clear mechanism so that every pixel can be reset to a well defined state.
The study of the clearing properties of new devices is therefore crucial.

\section{Acknowledgements}

The presented work has been carried out in collaboration with the Semiconductor Laboratory in Munich. The authors would
like to thank the DEPFET team for their support.

\newcommand{\NIM     }[3]{Nucl. Inst. Meth. {\bf #1} (#2) #3}
\newcommand{\IEEEJSSC}[4]{IEEE J. Solid-State Circuits, vol. #1, no. #2, p. #3, #4}

\end{document}